# Electric modulation of the Fermi arc spin transport via three-terminal configuration in the topological semimetal nanowires

Guang-Yu Zhu,[1,2,*] Ji-Ai Ning,[1,2,*] Jian-Kun Wang[1,2,*], Xin-Jie Liu,[1,2] Jing-Zhi Fang,[1,2] Ze-Nan Wu,[1,2] Jia-Jie Yang,[1,2] Ben-Chuan Lin,[1,2,3,†] Shuo Wang,[1,2,3,#] Dapeng Yu[1,2,3]

[1]*Shenzhen Institute for Quantum Science and Engineering, Southern University of Science and Technology, Shenzhen, 518055, China*

[2]*International Quantum Academy, Shenzhen 518048, China.*

[3]*Guangdong Provincial Key Laboratory of Quantum Science and Engineering, Southern University of Science and Technology, Shenzhen, 518055, China*

[*] *These authors contributed equally to this work.*
[†] linbc@sustech.edu.cn
[#] wangs6@sustech.edu.cn

**Abstract**

Spin momentum locking is a key feature of the topological surface state, which plays an important role in spintronics. The electrical detection of current-induced spin polarization protected by the spin momentum locking in non-magnetic systems provides a new platform for developing spintronics while previous studies were mostly based on magnetic materials. In this study, the spin transport measurement of Dirac semimetal $Cd_3As_2$ was studied by the three-terminal geometry, and a hysteresis loop signal with high resistance and low resistance state was observed. The hysteresis was reversed by reversing the current direction, which illustrates the spin-momentum locking feature of $Cd_3As_2$. Furthermore, we realized the on-off states of the spin signals through electric modulation of the Fermi arc via the three-terminal configuration, which enables the great potential of $Cd_3As_2$ in spin field-effect transistors.

Dirac semimetal has obtained great attention because of its unique quantum transport characteristics, such as higher carrier mobility[1-4], negative magnetoresistance caused by chiral anomaly[5-8] as well as extremely large magnetoresistance[3,9,10]. The Dirac semimetal can be transformed into the Weyl semimetal when the time-reversal symmetry is broken[11]. The surface dispersion relation of a Weyl semimetal is topologically equivalent to a non-compact Riemann surface without equal-energy contour that encloses the projection of the Weyl point[11], leading to the emergence of Fermi arcs, a unique topological surface state[12-17]. Fermi arcs have been observed directly by Angle-resolved photoemission spectroscopy, which has the characteristic of spin-momentum locking[18,19].

Spin-momentum locking is when the spin of electrons is locked perpendicular to the carrier momentum as shown in Fig. 1(a). It is caused by a strong spin-orbit interaction that has been observed earlier in the three-dimensional topological insulators[20,21] as well as the topological Dirac/Weyl semimetals[18,22-24] through optical and transport measurements. Since that current can induce spin polarization protected by the spin momentum locking, the electrical detection as well as controlling of this key feature is valuable for its applications in energy-efficient spintronic devices[22,25-27]. However, previous studies of the spin momentum locking property in topological materials including topological insulators, mostly reported the electric current control of the spin signal while few reported the electric gate control of the spin signal. The gate control is important, because if the spin signal can be turned on/off, the probable spintronics application would be much promising. For example, a novel type of spin field effect transistor could be realized.

In this work, the spin transport measurements of Dirac semimetal $Cd_3As_2$ nanowires were carried out using a three-terminal configuration[28]. The resistance demonstrates a hysteresis loop as the external magnetic fields sweep back and forth. When the magnetization M of the ferromagnetic electrode (Co) is parallel (anti-parallel) to the spin polarization *s* of the carriers, a high (low) voltage state is demonstrated. Such a hysteresis resistance state can be reversed by altering the direction of the current, revealing the spin-momentum locking feature of the $Cd_3As_2$ nanowires. Furthermore, the spin signal can be tuned by the gate voltage, enabling the spin transport on-off states.

The $Cd_3As_2$ nanowires were grown by chemical vapor deposition (CVD) having a high quality, large surface volume ratio, and low carrier density[24,29]. Three terminal geometry measurements were applied for the electrical detection of the spin transport of $Cd_3As_2$ nanowires. The schematic of the device is displayed in Fig. 1(b), in which the ferromagnetic electrode (Co) indicated by green is located between the two outer nonmagnetic electrodes (Au) indicated by yellow. The left Au electrode and the middle Co electrode were used to apply current, while the right Au electrode and the middle Co were used to measure the spin signal. The three-terminal geometry was used to separate the current and the voltage circuits, which can limit the anisotropic magnetoresistance[30] and anomalous Hall effect induced by the Co electrodes[31]. An external in-plane magnetic field *B* paralleled to the long axis of the Co electrode was applied to control the magnetization of the Co electrode. The direction along +y is defined as positive B while the current along +x is defined as positive. The contacts were patterned via electron beam lithography and were deposited sequentially by electron beam evaporation (EBE). The scanning electron microscopy (SEM) image of the fabricated device is shown in Fig. 1(c), in which the $Cd_3As_2$ nanowire has a diameter of about 150 nm, and the width of the ferromagnetic electrode (Co) is 500 nm while the width of the Au electrodes is 1μm.

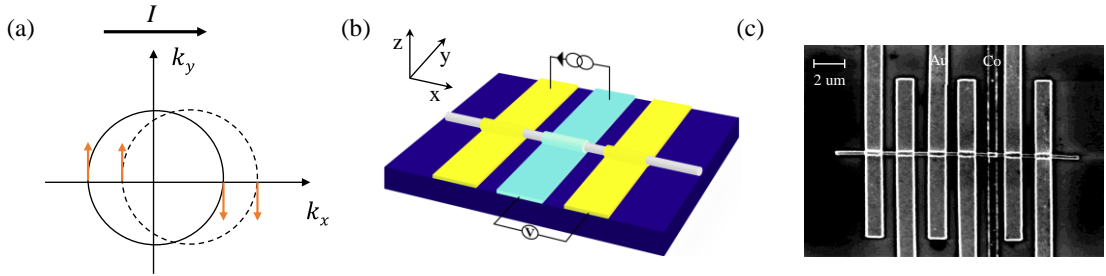

FIG. 1. (a) Schematic illustrations of the $k_x$-$k_y$ plane of the surface states. Spin is locked to the momentum at each point. (b) The three-terminal geometry measurement diagram, which used to separate the current and voltage circuits. The middle is a Co electrode indicated by blue and the two outer electrodes are Au electrodes indicated by yellow. (c) The scanning electron microscopy image of a typical device.

The three-terminal geometry measurement under T= 2 K was carried out and the results are shown in Fig. 2. In which a high voltage state and a low voltage state displayed as a hysteresis loop were observed as the external in-plane magnetic field was sweeping back and forth. When $I_{dc}$ = 100 nA, the momentum of carries is along the -x direction (indicated by the black arrow). Due to the spin-momentum locking, the left-moving electrons have a spin-up polarization *s* pointing along the +*y* direction (indicated by the red arrow) as shown in Fig. 2(a). When the magnetization of the Co electrode is +M (indicated by the green arrow) under a positive field, the relative orientation between *s* and +M is paralleled which leads to a high resistance state[32]. When the field wept from a positive to a negative field, the magnetization of the Co electrode is -M, therefore, *s* is anti-paralleled to the -M, and a low resistance state was formed.

When the current direction was reversed i.e., $I$ = -100 nA, the spin polarization of electrons *s* is also reversed due to the spin-momentum locking (indicated by the red arrow) as shown in Fig. 2(b). Therefore, the relative orientation between *s* and +M is anti-paralleled under a positive and paralleled under a negative field, respectively. As a result, a reversed hysteresis loop signal can be formed as shown in Fig.2 (b). The reversed current leading to the reversed hysteresis loop unambiguously reveals the spin momentum-locking properties of $Cd_3As_2$ nanowires.

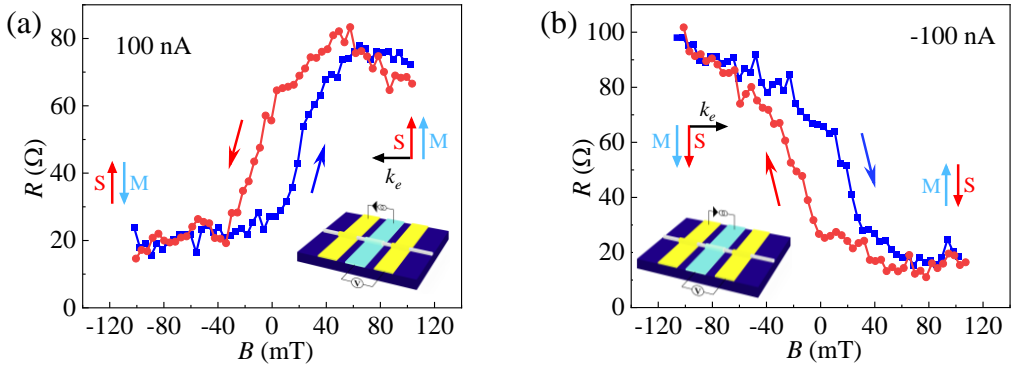

FIG. 2. Three-terminal measurement at 2 K. (a) A hysteretic loop signal measured under $I_{dc}$ = 100 nA. (b) A hysteretic loop signal measured under $I_{dc}$ = -100 nA. The high resistance states and low resistance states are reversed as the current direction was reversed, showing the spin-momentum locking feature.

Further, gate voltage $V_g$ tunable measurements were carried out, and the results are shown in Fig. 3(a). When the applied gate voltage is in the range from -4 V to +4 V, an obvious hysteresis loop signal can be observed. As the voltage increases, the hysteresis loop signal is weakened and when the voltage reached ±6 V, the hysteresis loops are vague. Eventually, the spin signals disappeared at a negative voltage of -8 V and a positive +10 V, respectively.

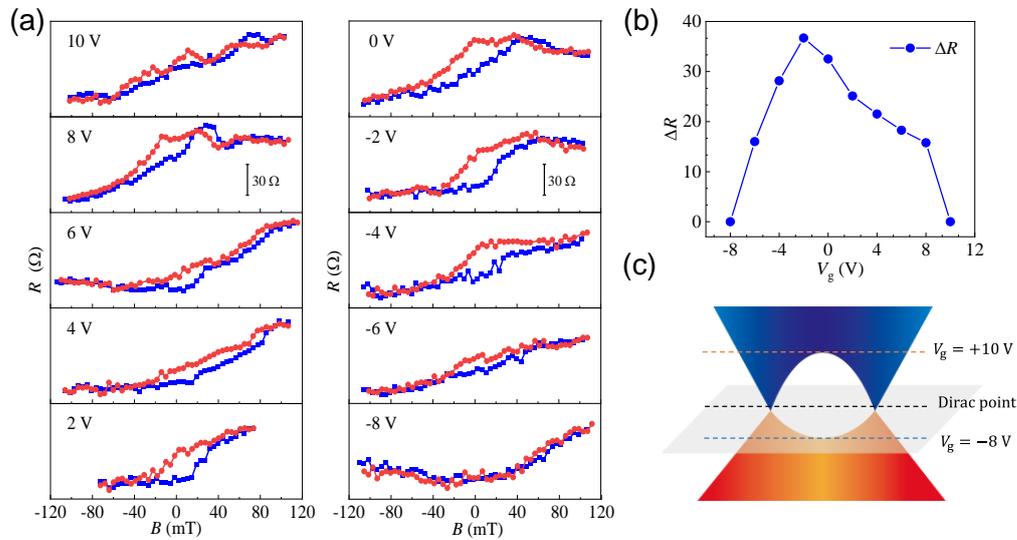

FIG. 3. (a) The hysteretic loop signals under various gate voltages. (b) The gate voltage $V_g$ dependence of the height between the high resistance state and the low resistance state. (c) Schematic diagram of the spin signal on-off state under various gate voltages.

To illustrate the gate dependence of the spin signals, we define the height between the high resistance and the low resistance states as $\Delta R$. The extracted ΔR from Fig. 3(a) as a function of the gate voltage is shown in Fig. 3(b). The values of ΔR display a peak at $V_g$ = -2 V, and as the absolute value of gate voltage increases, the ΔR decreases. The spin signals are significantly tuned by the gate voltage. Moreover, on-off states can be realized through modulate the gate voltage. The complete on-off spin transport states tuned by the gate voltage enable $Cd_3As_2$ have great potential in the spin field-effect transistor. The mechanism of the spin signal tuned by gate voltages can be understood following. When the Fermi level is closed to the Dirac point $V_g$ = -2 V, as shown in Fig. 3(c), the spin signal contributed by the Fermi arc reaches its maximum value. When the

Fermi level is far from the Dirac point, the contributions from the Fermi arc are weakened, resulting in a decrease in the spin signals. As the applied $V_g$ is larger than 10 V or $V_g$ is less than -8 V, the spin signals disappear completely. The reason for the complete disappearance of the signal may be attributed to the appearance of a topological phase transition, which is likely to come from the Lifshitz transition[9-15,17,24].

In summary, we observed a hysteresis loop signal with high resistance and low resistance state through three-terminal geometry measurements which reveal current-induced spin-polarization in the surface state of Dirac semimetals $Cd_3As_2$ nanowires. The three-terminal configuration could have much less channel length in the future spintronics application. The high resistance and low resistance state are reversed while reversing the current direction, illustrating the spin-momentum locking feature of $Cd_3As_2$ nanowires. Further, the on-off control of the spin signal can be realized by tuning the gate voltage, enabling the great potential of $Cd_3As_2$ nanowires in the spin field-effect transistor. Our results provide a new perspective for the future spintronic devices.

**Acknowledgements:** This work was supported by National Natural Science Foundation of China (No. 12004158, No. 12074162), National Key Research and Development Program of China (No. 2020YFA0309300), the Key-Area Research and Development Program of Guangdong Province (No. 2018B030327001) and Guangdong Provincial Key Laboratory (Grant No.2019B121203002).

**Declaration of competing interest**

The authors have no competing interests to declare that are relevant to the content of this article.

**Data availabitly**

The data that support the findings of this study are available from the corresponding author upon reasonable request.